# MAGNETIC SYMMETRY OF THE PLAIN DOMAIN WALLS IN THE PLATES OF CUBIC FERRO- AND FERRIMAGNETS

B. M. Tanygin[1], O. V. Tychko[2]

Kyiv Taras Shevchenko National University, Radiophysics Faculty, Volodumiurska 64, Kyiv, MSP 01601, Ukraine

[1]E-mail: b.m.tanygin@gmail.com, [2]E-mail: a.tychko@gmail.com

**Abstract.** Magnetic symmetry of possible plane domain walls in arbitrary oriented plates of the crystal of hexoctahedral crystallographic class is considered. The symmetry classification is applied for ferro- and ferrimagnets.

## 1. Introduction

For sequential examination of static and dynamic properties of domain walls (DWs) in magnetically ordered media it is necessary to take into account their magnetic symmetry [1,2]. The complete symmetry classification of plane 180°-DWs in magnetically ordered crystals [1], similar classification of these DWs with Bloch lines in ferromagnets and ferrites [2] and magnetic symmetry classification of plane non $180^0$-DWs (all possible DW types including $0^0$-DWs [3]) in ferro- and ferrimagnets [4] were carried out earlier. These DW symmetry classifications allows arbitrary crystallographic point symmetry group of the crystal. The influence of the spatially restricted sample surfaces on the DW magnetic symmetry wasn't considered in works [1-4]. The real magnetic sample restricts a spatial (3D) magnetization distribution. Therefore, it modifies the DW symmetry in general case. This paper presents the investigation of the influence of the restricted sample surfaces on the symmetry of the all possible ($0^0$-, $60^0$-, $70.5^0$-, $90^0$-, $109.5^0$-, $120^0$- and $180^0$-DW [4,5]) plane (i.e. DW with $r_0 \gg \delta$, where $r_0$ is the curvature radius of the DW [1]) DWs in an arbitrary oriented plate of the cubic (crystallographic point symmetry group m3m) ferro- and ferrimagnets.

## 2. Domain wall symmetry in the restricted sample

The DW symmetry can be described by the magnetic symmetry classes (MSCs) $G_k$ where $k$ is a MSC number [1]. The MSC $G_k$ of DW is the magnetic symmetry group including all symmetry transformations (all translations are considered as unit operations) that do not change the spatial distribution of magnetic moments in the crystal with DW. The above-mentioned group is a subgroup of the magnetic (Shubnikov's) symmetry group $G_P^\infty$ of the crystal paramagnetic phase [6]. Total number of MSCs of arbitrary type DWs (i.e. DWs with arbitrary $2\alpha$ angle ($0^0 \leq 2\alpha \leq 180^0$) between the unit time-odd axial vectors $m_1$ and $m_2$ directed along magnetization vectors $M_1$ and $M_2$ in neighboring domains) in ferro- and ferrimagnets is equal to 64. General enumeration of MSCs contains 42 MSCs ($1 \leq k \leq 42$) of 180°-DWs [1], 10 MSCs ($7 \leq k \leq 13$ and $16 \leq k \leq 18$) of non 0°/180°- DWs and 42 MSCs ($k=2$, $6 \leq k \leq 13$, $16 \leq k \leq 19$, $k=22$, 24, 26, 30, 32, 37, 39 and $43 \leq k \leq 64$) of 0°-DWs [4]. The MSCs ($k=25$, 28, 37-41, 52, 54, 61-63) with six-fold symmetry axes (including inversion axes) do not realized in the cubic crystals [4].

The unified co-ordinate system $O\tilde{x}\tilde{y}\tilde{z}$ is chosen as $[e_{\tilde{x}}, e_{\tilde{y}}, e_{\tilde{z}}] = [a_2, -a_1, n_W]$ where $n_W$ is the unit polar time-even vector along the DW plane normal [4]. For the 180°-DWs the vectors $a_1$ and $a_2$ are given early [1] as vectors $\tau_1$ and $\tau_2$ respectively. For the case of $2\alpha \neq 180°$ the unit vector $a_1$ coincides with the direction of the vector $\Delta m - n_W(n_W \Delta m)$ (at $b_\Delta \neq 0$ and $b_\Sigma = 0$) or $[a_2 \times n_W]$ (at $b_\Delta = 0$ or $b_\Sigma \neq 0$) where $\Delta m = m_2 - m_1$, $b_\Delta = |[n_W \times \Delta m]|$, $b_\Sigma = |[n_W \times m_\Sigma]|$. Here the unit vector $a_2$ coincides with the direction of vector $m_\Sigma - n_W(n_W m_\Sigma)$ (at $b_\Sigma \neq 0$) or $[n_W \times a_1]$ (at $b_\Delta \neq 0$ and $b_\Sigma = 0$) or else with an arbitrary direction in the DW plane ($a_2 \perp n_W$ at $b_\Sigma = b_\Delta = 0$) where $m_\Sigma = m_1 + m_2$. The mutual orientation of the vectors $m_1$, $m_2$ and $n_W$ is determined by the parameters $a_\Sigma = (n_W m_\Sigma)$, $a_\Delta = (n_W \Delta m)$, $a_C = (n_W m_C)$, $b_\Sigma$ and $b_\Delta$, where $m_C = [m_1 \times m_2]$. The mutual orientation of the vectors $m_1$, $m_2$, $n_W$ and $n_S$ is determined by parameters $a_1 = (a_1 n_S)$, $a_2 = (a_2 n_S)$, $a_n = (n_W n_S)$, $b_1 = |[n_S \times a_1]|$, $b_2 = |[n_S \times a_2]|$ and $b_n = |[n_S \times n_W]|$, where $n_S$ is sample plane normal.

The MSC $G_P$ of restricted sample of crystal in paramagnetic phase could be defined as $G_P = G_P^\infty \cap G_S$ where sample shape MSC $G_S$ is $\infty/mmm1'$ for volumetric plate. MSCs of DWs in volumetric plate should satisfy the condition $G_k \subset G_P$. The MSCs of the all possible plane DWs in the arbitrary oriented plate of cubic crystals of hexoctahedral class (crystallographic point symmetry group $m3m$ in the paramagnetic phase [6]) are presented in table. Here symmetry axes are collinear with vectors $a_1$ and $a_2$ and reflection planes are perpendicular to them. For MSCs with $k$ =24, $k$ =26, $k$ =27, 29$\leq k \leq$36, 42$\leq k \leq$51, $k$ =53, 55$\leq k \leq$60 and $k$=64 only generative symmetry elements are represented.

General enumeration of MSCs of 0°-DWs contains MSCs with $k$=2, $6 \leq k \leq 13$, $16 \leq k \leq 19$, $k$=22, 24, 26, 30, 32, $43 \leq k \leq 51$, $k$=53, $55 \leq k \leq 60$ and $k$=64. The $60^0$- and 120°-DWs are represented by MSCs with $k$=10, 16, 18 and $k$=11, 13, 16 respectively. The MSCs of the $70.5^0$-, $90^0$- (both for <100> and <110> like easy magnetization axis [5]) and $109.5^0$-DWs are the MSCs with 7<$k$<13, 16<$k$<18. The general list of 180°-DWs includes MSCs with $1 \leq k \leq 42$ except for $k$=25, 28, 37-41.

### 3. Conclusions

The complete collection of (*nml*)-plates with all possible orientations includes the full list of MSCs of $2\alpha$-DWs in cubic $m3m$ crystal. For separate (*nml*)-plates with fixed combination of Miller indexes this list is limited. Such limitation depends on plate orientation. It is minimal and maximal for the samples with high-symmetry (such as (100)-, (110)- or (111)-plates) and low-symmetry (the (*nml*)-plates, where indexes are non-zero and have different absolute values) developed surface respectively. Maximal quantity of MSCs of $2\alpha$-DWs is for (100)-plates. The MSC with $k=16$ is the MSC of all above-mentioned $2\alpha$-DWs in arbitrary oriented plate of cubic $m3m$ crystal.

**Table.** MSCs of the plane $2\alpha$-DWs in plates of the cubic m3m crystal.

| k | {nml}-sample | $m_1$, $m_2$, $n_W$ and $n_S$: mutual orientation**. | $m_1$, $m_2$ and $n_W$: mutual orientation | Symmetry elements*** | MSC symbol |
|---|---|---|---|---|---|
| 1 | {100}, {110} | $b_n = 0$ or $b_1 = 0$ | $a_\Sigma = b_\Sigma = a_\Delta = 0$ | $(1, 2_2, \bar{2}_1, \bar{2}_n) \times (1, \bar{1})$ | mmm |
| 2 | {100}, {110} | $b_n = 0$ or $b_1 = 0$ | $a_\Delta = b_\Delta = a_\Sigma = 0$ or $a_\Sigma = b_\Sigma = a_\Delta = 0$ | $1, \bar{2}'_1, \bar{2}_2, 2'_n$ | mm′2′ |
| 3 | {100}, {110} | $b_n = 0$ or $b_1 = 0$ | $a_\Sigma = b_\Sigma = a_\Delta = 0$ | $1, 2_2, \bar{2}_1, \bar{2}_n$ | mm2 |
| 4 | {nml}* | $a_1 = 0$ or $b_1 = 0$ | $a_\Sigma = b_\Sigma = a_\Delta = 0$ | $1, \bar{1}', 2'_1, \bar{2}_1$ | 2′/m |
| 5 | {nml}* | $a_n = 0$ or $b_n = 0$ | $a_\Sigma = b_\Sigma = a_\Delta = 0$ | $1, \bar{1}', 2'_n, \bar{2}_n$ | 2′/m |
| 6 | {nml}* | $a_2 = 0$ or $b_2 = 0$ | $a_\Sigma = a_\Delta = a_C = 0$ | $1, \bar{2}_2$ | m |
| 7 | {100}, {110} | $b_n = 0$ or $b_1 = 0$ | $a_\Sigma = a_\Delta = 0$ | $1, 2'_1, 2_2, 2'_n$ | 2 2′ 2′ |
| 8 | {nml}* | $a_n = 0$ or $b_n = 0$ | $a_\Sigma = a_\Delta = 0$ | $1, 2'_n$ | 2′ |
| 9 | {100}, {110} | $b_n = 0$ or $b_1 = 0$ | $a_C = a_\Delta = b_\Sigma = 0$ | $1, 2'_1, \bar{2}'_2, \bar{2}_n$ | m m′ 2′ |
| 10 | {nml}* | $a_1 = 0$ or $b_1 = 0$ | $a_\Delta = 0$ | $1, 2'_1$ | 2′ |
| 11 | {nml}* | $a_n = 0$ or $b_n = 0$ | $a_C = a_\Delta = b_\Sigma = 0$ | $1, \bar{2}_n$ | m |
| 12 | {nml}* | $a_1 = 0$ or $b_1 = 0$ | $a_C = 0$ | $1, \bar{2}'_1$ | m′ |
| 13 | {nml}* | $a_2 = 0$ or $b_2 = 0$ | $a_\Sigma = 0$ | $1, 2_2$ | 2 |
| 14 | {nml}* | $a_2 = 0$ or $b_2 = 0$ | $a_\Sigma = b_\Sigma = 0$ | $1, \bar{1}, 2'_2, \bar{2}'_2$ | 2′/m′ |
| 15 | Arbitrary | Arbitrary | $a_\Sigma = b_\Sigma = 0$ | $1, \bar{1}'$ | $\bar{1}'$ |
| 16 | Arbitrary | Arbitrary | Arbitrary | 1 | 1 |
| 17 | {100}, {110} | $b_n = 0$ or $b_1 = 0$ | $a_C = a_\Sigma = b_\Delta = 0$ | $1, \bar{2}'_1, 2_2, \bar{2}'_n$ | m′ m′ 2 |
| 18 | {nml}* | $a_n = 0$ or $b_n = 0$ | $a_C = a_\Sigma = b_\Delta = 0$ | $1, \bar{2}'_n$ | m′ |
| 19 | {nml}* | $a_n = 0$ or $b_n = 0$ | $b_\Delta = b_\Sigma = 0$ | $1, 2_n$ | 2 |
| 20 | {nml}* | $a_n = 0$ or $b_n = 0$ | $a_\Sigma = b_\Sigma = b_\Delta = 0$ | $1, \bar{1}', 2_n, \bar{2}'_n$ | 2/m |
| 21 | {100}, {110} | $b_n = 0$ or $b_1 = 0$ | $a_\Sigma = b_\Sigma = b_\Delta = 0$ | $1, 2_1, 2_2, 2_n$ | 222 |
| 22 | {100}, {110} | $b_n = 0$ or $b_1 = 0$ | $b_\Delta = b_\Sigma = 0$ | $1, \bar{2}'_1, \bar{2}'_2, 2_n$ | m′ m′ 2 |
| 23 | {100}, {110} | $b_n = 0$ or $b_1 = 0$ | $a_\Sigma = b_\Sigma = b_\Delta = 0$ | $(1, 2_1, 2_2, 2_n) \times (1, \bar{1}')$ | m′m′m′ |
| 24 | {111} | $b_n = 0$ | $b_\Delta = b_\Sigma = 0$ | $3_n$ | 3 |
| 26 | {111} | $b_n = 0$ | $b_\Delta = b_\Sigma = 0$ | $3_n, \bar{2}'_1$ | 3m′ |
| 27 | {111} | $b_n = 0$ | $a_\Sigma = b_\Sigma = b_\Delta = 0$ | $3_n, 2_1$ | 32 |
| 29 | {111} | $b_n = 0$ | $a_\Sigma = b_\Sigma = b_\Delta = 0$ | $\bar{3}'_n, \bar{2}'_1$ | $\bar{3}$′m′ |
| 30 | {100} | $b_n = 0$ | $b_\Delta = b_\Sigma = 0$ | $4_n$ | 4 |
| 31 | {100} | $b_n = 0$ | $a_\Sigma = b_\Sigma = b_\Delta = 0$ | $4_n, \bar{2}'_n$ | 4/m′ |
| 32 | {100} | $b_n = 0$ | $b_\Delta = b_\Sigma = 0$ | $4_n, \bar{2}'_1$ | 4m′m′ |
| 33 | {100} | $b_n = 0$ | $a_\Sigma = b_\Sigma = b_\Delta = 0$ | $4_n, 2_1$ | 422 |
| 34 | {100} | $b_n = 0$ | $a_\Sigma = b_\Sigma = b_\Delta = 0$ | $4_n, \bar{2}'_1, \bar{2}'_n$ | 4/m′m′m′ |
| 35 | {100} | $b_n = 0$ | $a_\Sigma = b_\Sigma = b_\Delta = 0$ | $\bar{4}'_n$ | $\bar{4}$′ |
| 36 | {100} | $b_n = 0$ | $a_\Sigma = b_\Sigma = b_\Delta = 0$ | $\bar{4}'_n, 2_1$ | $\bar{4}$′2m′ |
| 42 | {111} | $b_n = 0$ | $a_\Sigma = b_\Sigma = b_\Delta = 0$ | $\bar{3}'_n$ | $\bar{3}$′ |

**Table.** MSCs of the plane $2\alpha$-DWs in plates of the cubic m3m crystal (continue).

| k | $\{nml\}$-sample | $m_1$, $m_2$, $n_W$ and $n_S$: mutual orientation. | $m_1$, $m_2$ and $n_W$: mutual orientation | Symmetry elements | MSC symbol |
|---|---|---|---|---|---|
| 43 | $\{100\}$, $\{110\}$ | $b_n = 0$ or $b_l = 0$ | $a_\Delta = b_\Delta = a_\Sigma = 0$ | $(1, 2'_1, \bar{2}_2, \bar{2}'_n) \times (1, \bar{1})$ | mm'm' |
| 44 | $\{100\}$, $\{110\}$ | $b_n = 0$ or $b_l = 0$ | $a_\Delta = b_\Delta = a_\Sigma = 0$ | $1, 2'_1, \bar{2}_2, \bar{2}'_n$ | mm' 2' |
| 45 | $\{nml\}$* | $a_2 = 0$ or $b_2 = 0$ | $a_\Delta = b_\Delta = a_\Sigma = 0$ | $1, \bar{1}, 2_2, \bar{2}_2$ | 2/m |
| 46 | $\{nml\}$* | $a_n = 0$ or $b_n = 0$ | $a_\Delta = b_\Delta = a_\Sigma = 0$ | $1, \bar{1}, 2'_n, \bar{2}'_n$ | 2'/m' |
| 47 | $\{nml\}$* | $a_1 = 0$ or $b_1 = 0$ | $a_\Delta = b_\Delta = 0$ | $1, \bar{1}, 2'_1, \bar{2}'_1$ | 2'/m' |
| 48 | Arbitrary | Arbitrary | $a_\Delta = b_\Delta = 0$ | $1, \bar{1}$ | $\bar{1}$ |
| 49 | $\{nml\}$* | $a_n = 0$ or $b_n = 0$ | $a_\Delta = b_\Delta = b_\Sigma = 0$ | $1, \bar{1}, 2_n, \bar{2}_n$ | 2/m |
| 50 | $\{100\}$, $\{110\}$ | $b_n = 0$ or $b_l = 0$ | $a_\Delta = b_\Delta = b_\Sigma = 0$ | $1, 2'_1, 2'_2, 2_n$ | 2 2' 2' |
| 51 | $\{100\}$, $\{110\}$ | $b_n = 0$ or $b_l = 0$ | $a_\Delta = b_\Delta = b_\Sigma = 0$ | $(1, 2'_1, 2'_2, 2_n) \times (1, \bar{1})$ | mm'm' |
| 53 | $\{111\}$ | $b_n = 0$ | $a_\Delta = b_\Delta = b_\Sigma = 0$ | $3_n, 2'_1$ | 32' |
| 55 | $\{111\}$ | $b_n = 0$ | $a_\Delta = b_\Delta = b_\Sigma = 0$ | $\bar{3}_n, 2'_1$ | $\bar{3}$ m' |
| 56 | $\{100\}$ | $b_n = 0$ | $a_\Delta = b_\Delta = b_\Sigma = 0$ | $4_n, \bar{2}_n$ | 4/m |
| 57 | $\{100\}$ | $b_n = 0$ | $a_\Delta = b_\Delta = b_\Sigma = 0$ | $4_n, 2'_1$ | 42'2' |
| 58 | $\{100\}$ | $b_n = 0$ | $a_\Delta = b_\Delta = b_\Sigma = 0$ | $4_n, \bar{2}'_1, \bar{2}_n$ | 4/mm'm' |
| 59 | $\{100\}$ | $b_n = 0$ | $a_\Delta = b_\Delta = b_\Sigma = 0$ | $\bar{4}_n$ | $\bar{4}$ |
| 60 | $\{100\}$ | $b_n = 0$ | $a_\Delta = b_\Delta = b_\Sigma = 0$ | $\bar{4}_n, 2'_1$ | $\bar{4}$2'm' |
| 64 | $\{111\}$ | $b_n = 0$ | $a_\Delta = b_\Delta = b_\Sigma = 0$ | $\bar{3}_n$ | $\bar{3}$ |

\* ($nml$)-plates with arbitrary Miller indexes except non zero values $|n| \neq |m| \neq |l| \neq |n|$

\*\* At $(n_W a_1) = (n_W a_2) = 0$

\*\*\* The possible symmetry elements are rotations around two-fold symmetry axes $2_n$, $2'_n$ or $2_1$, $2'_1$ or else $2_2$, $2'_2$ that are collinear with the unit vectors $n_W$ or $a_1$ or else $a_2$, respectively, reflections in planes $\bar{2}_n$, $\bar{2}'_n$ or $\bar{2}_1$, $\bar{2}'_1$ or else $\bar{2}_2$, $\bar{2}'_2$ that are normal to the above mentioned vectors, respectively, rotations around three-, four-fold symmetry axes $3_n$, $4_n$ that are collinear with the vector $n_W$, rotations around three-, four-fold inversion symmetry axes $\bar{3}_n, \bar{3}'_n, \bar{4}_n, \bar{4}'_n$ that are collinear with the vector $n_W$, inversion in the symmetry center $\bar{1}$, $\bar{1}'$ and identity 1. Here an accent at symmetry elements means a simultaneous use of the time reversal operation [6].